\documentclass[12pt,reqno,fleqn,a4paper]{amsart}
\usepackage{mathrsfs}
\usepackage{bbm}
\usepackage{amsmath,amssymb}
\usepackage{color}

\usepackage[titletoc]{appendix}

\allowdisplaybreaks \numberwithin{equation}{section}
\numberwithin{equation}{section} 

 \newtheorem{thm}{Theorem}[section]
 
 \newtheorem{lem}[thm]{Lemma}
 \newtheorem{prp}[thm]{Proposition}

\newenvironment{prf}{\noindent {\it Proof} \ }{\hfill $\Box$}

\newcommand{\eqa}{\begin{eqnarray}}
\newcommand{\eeqa}{\end{eqnarray}}

\newcommand{\beq}{\begin{equation}}
\newcommand{\eeq}{\end{equation}}

\newcommand\pd{\partial} \newcommand\od{\mathrm{d}}

\newcommand{\nn}{\nonumber}

\newcommand{\Gm}{\Gamma}

\newcommand{\dt}{\delta}

\newcommand\C{\mathbb{C}}

\newcommand\Z{\mathbb{Z}}
\newcommand\Zop{\mathbb{Z^{\mathrm{odd}}_+}}

\newcommand\cA{\mathcal{A}}
\newcommand\B{\mathcal{B}}

\newcommand\cG{\mathcal{G}}
\newcommand\cL{\mathcal{L}}
\newcommand\cM{\mathcal{M}}

\begin{document}
\title{Block algebra in two-component BKP and D type Drinfeld-Sokolov hierarchies}
\author{
Chuanzhong Li\dag,\  \ Jingsong He\ddag}
\allowdisplaybreaks
\dedicatory {\small Department of Mathematics,  Ningbo University, Ningbo, 315211, China\\
\dag lichuanzhong@nbu.edu.cn\\
\ddag hejingsong@nbu.edu.cn}
\thanks{\ddag Corresponding author}

\date{}

\begin{abstract}
We construct generalized additional symmetries of a two-component
BKP hierarchy defined by two pseudo-differential Lax operators.
These additional symmetry flows form a Block type algebra with some
modified(or additional) terms because of a B type reduction
condition of this integrable hierarchy. Further we  show that the D
type Drinfeld-Sokolov hierarchy, which is a reduction of the
two-component  BKP hierarchy, possess a complete Block type
additional symmetry algebra. That  D type Drinfeld-Sokolov hierarchy
has a  similar algebraic structure  as the bigraded Toda hierarchy
which is a differential-discrete integrable system.
\end{abstract}

\allowdisplaybreaks
\maketitle{\allowdisplaybreaks}

\allowdisplaybreaks
 \setcounter{section}{0}

\section{Introduction}

One interesting  topic in the study of  integrable hierarchies  is
to find symmetry and its recursion relation,and further to identify
its algebraic structures. There are already many results in
literatures, for example \cite{fokasstudy}-\cite{Mikhailov2}. Among
these symmetries, the additional symmetry is a relatively new type
and has been studied extensively in recent years, which contains
dynamic variables explicitly and does not commutes with each other.
Additional symmetries of the Kadomtsev-Petviashvili(KP) hierarchy
was introduced  by Orlov and Shulman \cite{os1} which contain one
kind of important symmetry called Virasoro symmetry.  These
symmetries form a centerless $W_{1+\infty}$ algebra is closely
related to matrix model by means of the Virasoro constraint and
string equation\cite{D witten,Douglas, dl1,asv1,asv2}. Two
sub-hierarchies of KP, BKP hierarchy and CKP
hierarchy\cite{DKJM-KPBKP,DJKM}, have been shown to possess
additional symmetry\cite{kt1}-\cite{heLMP} with consideration of
the reductions on the Lax operators.

The 2-dimensional Toda Lattice(2dTL) hierarchy is introduced by Ueno
and Takasaki in \cite{takasaki1} based on the Sato theory. It is
natural to construct the additional symmetry of the 2dTL hierarchy
because of the similarity between the KP hierarchy and 2dTL
hierarchy\cite{asv2}. For the dispersionless Toda
hiearchy\cite{Takasakicmp,Takasaki},  additional symmetry is used to give
string equations and Riemann-Hilbert problem. Note that, there exist
two different sub-hierarchies of 2dTL hierarchy, 2-dimensional B type Toda Lattice (2dBTL) and
2-dimensional C type Toda Lattice (2dCTL) \cite{takasaki1} which correspond to the infinite-dimensional
algebras $w_{\infty}^B\times w_{\infty}^B$ and $w_{\infty}^C\times w_{\infty}^C$. The additional symmetry of
the 2dBTL and 2dCTL hierarchies have been given recently in
\cite{chenghe2}. These results show that additional symmetry is
one kind of general features of the integrable hierarchies.

As a generalization of Virasoro algebra, Block type
infinite-dimensional Lie algebra and its representation theory  have
been studied intensively in references\cite{Block}-\cite{Su}. The Block type Lie algebra $\bar B$ without cental extension is defined as
\begin{eqnarray}\bar B=span\{L_{m,l},\ m,l\in \Z,l\geq 0\},\end{eqnarray}
with bracket
\begin{eqnarray}
[L_{m,l},L_{n,k}]=(mk-nl)L_{m+n-1,l+k-1}.
\end{eqnarray}
 Note
that the Virasoro algebra is one kind of widely used
infinite-dimensional algebra in mathematical physics,
particularly in integrable systems\cite{goddard}. However, it is
curious to note that, in the past 50 years after the introduction of
the Block algebra, there does not exist a result on the application
of this algebra in integrable systems until last year, to the best
of our knowledge. In paper\cite{ourBlock}, we provide  a novel Block
type algebraic structure  of the bigraded Toda hierarchy(BTH) with
the help of the additional symmetry. This is the first time to find
the direct relation between integrable hierarchy and the Block type
algebra. Here BTH as a general reduction of 2dTL hierarchy, is
introduced in \cite{C,TH} from the background of the topological
field and Gromov-Witten invariants.  The Hirota bilinear equations
and  solutions of the BTH are given in \cite{ourJMP,solutionBTH}.
Later Block algebra is found again in dispersionless bigraded Toda
 hierarchy \cite{dispBTH}, in two-dimensional Toda hierarchy\cite{Jipeng}.
Very recently Block algebra has been shown to have a close relation with
 3-algebra\cite{3algebra}.

Based on the above results of Block algebra, in order to explore the
universality of the Block type algebra in integrable systems, it is
necessary to find this kind of algebraic structure in the KP type
differential systems, due to the importance of the KP systems. In
\cite{kodama W algebra}, one kind of additional symmetry of the KP
hierarchy composed one generalized  $W_{1+\infty}$ algebra with
complicated structure coefficients but it was not a Block algebra.
Taking into consideration the complexity of the relevant formula of
the Block algebra,  as well as plenty of different possible
extensions of the KP hierarchy, it is a challenging problem to find
this algebra in the KP type hierarchies.

 It is a direct idea to consider the multi-component KP hierarchy(mcKP)\cite{sato1,jimbo1,takasakisigma,kaccomponent},
which has a sole Lax operator with matrix coefficients. However,
the algebra structure of the additional symmetry of the mcKP is very complicated and belongs
to the Virasoro type\cite{dickey1}. Note that integrable systems possessing the symmetry of
the Block type need to have two independent hierarchies of flows defined by two different
 Lax operators\cite{ourBlock,dispBTH,Jipeng}. But the flows of the  mcKP hierarchy with
 two pseudo-differential Lax operators are not well defined. Fortunately,
 for a two-component BKP hierarchy\cite{DJKM,shiota}, two Lax operators has been constructed in \cite{LWZ}(see eq.(3.3) and eq.(3.9) of this reference) from the view of  the Drinfeld-Sokolov Hierarchies of D Type.
 The Hamiltonian structure of this two-component BKP hierarchy is given in \cite{WX}.
 Therefore this two component BKP hierarchy \cite{LWZ} is a good candidate for us to
 explore the Block algebra in integrable hierarchy of KP type. In the following
text of this paper we construct the generalized additional symmetries of the two-component BKP hierarchy
 and identify its algebraic structure by using a similar method in \cite{Jipeng,CPAM}.
 Besides, the D type Drinfeld-Sokolov hierarchy is found to be a good differential model to derive complete
 Block type infinite dimensional Lie algebra.

This paper is arranged as follows. In next section we recall some
necessary facts of the two-component BKP hierarchy. In Sections 3,
we will give the generalized additional symmetries for the
two-component BKP hierarchy. By reducing the two-component BKP
hierarchy to the D type Drinfeld-Sokolov hierarchy, some concepts
and results about this reduced hierarchy will be introduced in
Section 4. The Block symmetries of Drinfeld-Sokolov hierarchy of
type D  will be derived in Section 5.
\section{Two component BKP hierarchy}

Let us firstly recall some basic facts\cite{LWZ,bkpds} of the two-component BKP hierarchy which
is well defined by two Lax operators.

 $\cA$ is assumed as an algebra of smooth functions of a spatial
coordinate $x$ and  derivation denoted as  $ D=\od/\od x$. This algebra  $\cA$ has following multiplying rule
\begin{equation*}
 D^i\cdot f =\sum_{r\geq0}\binom{i}{r}\, D^r(f)\, D^{i-r},
\quad f\in\cA.
\end{equation*}

For any operator $A=\sum_{i\in\Z} f_i D^i\in\cA$, its
nonnegative projection, negative projection, adjoint operator are
respectively defined as
\begin{align}\label{Apm}
&A_+=\sum_{i\geq0} f_i  D^i, \quad A_-=\sum_{i<0} f_i  D^i, \quad A^*=\sum_{i\in\Z}(- D)^i\cdot f_i.
\end{align}

Basing on definition  in\cite{LWZ}, the two Lax operators of the two-component BKP hierarchy have  form
\begin{equation} \label{PhP}
L= D+\sum_{i\ge1}u_i  D^{-i}, \quad \hat{L}=
D^{-1}\hat{u}_{-1}+\sum_{i\ge1}\hat{u}_i D^i,
\end{equation}
 such that
 \eqa \label{Bcondition}L^*=- D L D^{-1}\ \  \ \hat{L}^*=- D\hat{L}
D^{-1},\ \ \ r\in\Z_+.\eeqa
We call eq.\eqref{Bcondition} the B type condition of two-component BKP hierarchy.

The two-component BKP hierarchy is defined by the following
Lax equations:
\begin{align}\label{bkpLax}
& \frac{\pd \bar L}{\pd t_k}=[(L^k)_+, \bar L], \quad  \quad \frac{\pd \bar {L}}{\pd
\hat{t}_k}=[-(\hat{L}^k)_-, \bar {L}]
\end{align}
with $\bar L=L\ \  or\  \  \hat L,\ \ k\in\Zop$.

Note that $\pd/\pd t_1$ flow is equivalent to $\pd/\pd x$ flow, therefore it is reasonable to
assume $t_1=x$ in the following several sections.

One can write the operators $L$ and $\hat{L}$ in a dressing form as
\begin{equation} \label{PPh}
L=\Phi D\Phi^{-1},\quad \hat{L}=\hat{\Phi} D^{-1}\hat{\Phi}^{-1},
\end{equation}
where
\begin{align} \label{dreop}
\Phi=1+\sum_{i\ge 1}a_i D^{-i},\quad \hat{\Phi}=1+\sum_{i\ge 1}b_i
D^{i}
\end{align}
 satisfy
\begin{equation}\label{phipsi}
\Phi^*= D\Phi^{-1} D^{-1},\quad \hat{\Phi}^*= D\hat{\Phi}^{-1}
D^{-1}.
\end{equation}
Given $L$ and $\hat{L}$, the dressing operators $\Phi$ and
$\hat{\Phi}$ are determined uniquely up to a multiplication to the
right by operators with
constant coefficients. The two-component BKP hierarchy \eqref{bkpLax}
can also be redefined as
\begin{align}
&\frac{\pd \Phi}{\pd t_k}=- (L^k)_-\Phi, \quad
\frac{\pd \hat{\Phi}}{\pd t_k}=\bigl((L^k)_+ -\dt_{k1} \hat{L}^{-1}\bigr)\hat{\Phi}, \label{ppt1}\\
&\frac{\pd \Phi}{\pd \hat{t}_k}=- (\hat{L}^k)_-\Phi, \quad \frac{\pd
\hat{\Phi}}{\pd \hat{t}_k}=(\hat{L}^k)_+\hat{\Phi}, \label{ppt2}
\end{align}
with $k\in\Zop$.

Denote $t=(t_1,t_3,t_5,\dots)$,
$\hat{t}=(\hat{t}_1,\hat{t}_3,\hat{t}_5,\dots)$ and introduce
two wave functions
\begin{align}\label{wavef}
w(z)=w(t, \hat{t}; z)=\Phi e^{\xi(t;z)},
\quad \hat{w}(z)=\hat{w}(t, \hat{t}; z)=\hat{\Phi}
e^{x z+\xi(\hat{t};-z^{-1})},
\end{align}
where the function $\xi$ is defined as $\xi(t;
z)=\sum_{k\in\Zop} t_k z^k$. It is easy to see
$D^i e^{x z}=z^i e^{x z},\ \ i\in\Z$
and
\[
L\,w(z)=z w(z), \quad \hat{L} \hat{w}(z) = z^{-1} \hat{w}(z).
\]

The  two-component BKP hierarchy was proved to have infinitely many bi-Hamiltonian
structures and Hamiltonian densities which are the residues of $L^k$ and $\hat{L}^k$ with
tau-symmetric condition\cite{WX}.
The tau function of the  two-component BKP hierarchy  can be defined in form of the wave functions as
\begin{align}\label{wtau}
w(t,\hat{t};z)=\frac{\tau(t-2[z^{-1}],
\hat{t})}{\tau(t,\hat{t})}
e^{\xi(t;z)}, \quad \hat{w}(t,\hat{t};z)
=\frac{\tau(t,\hat{t}+2[z])}{\tau(t,\hat{t})}
e^{\xi(\hat{t};-z^{-1})}
\end{align}
where $[z]=\left(z,z^3/3,z^5/5,\dots\right)$.

With above preparation, it is time to  construct generalized additional symmetries for the two-component BKP hierarchy in the next section.

\section{Generalized additional symmetries of the two-component BKP hierarchy}

In this section, we are to construct generalized additional symmetries for the two-component BKP hierarchy by using the Orlov--Schulman operators whose coefficients
depend explicitly on the time variables of the
hierarchy.

With the same dressing operators given in eq.\eqref{dreop},
Orlov--Schulman operators $M,\hat M$ are constructed in following
dressing structure \cite{os1,bkpds}
\begin{equation*}\label{}
M=\Phi\Gm\Phi^{-1}, \quad
\hat{M}=\hat{\Phi}\hat{\Gm}\hat{\Phi}^{-1},
\end{equation*}
where
\[
\Gm=\sum_{k\in\Zop}k t_k  D^{k-1}, \quad \hat{\Gm}=x+\sum_{k\in\Zop}
k\hat{t}_k D^{-k-1}.
\]

Then it is easy to get  the following lemma.
\begin{lem}\label{thm-Mw}
The operators $M$ and $\hat{M}$ satisfy

\begin{equation}
[L, M]=1, \quad [\hat{L}^{-1},\hat{M}]=1;\ \
M w(z)=\pd_z w(z), \quad \hat{M} \hat{w}(z)=\pd_z \hat{w}(z);
\end{equation}
and for $\bar{M}=M$ or $\hat{M}$,
\begin{equation}\label{bkpMt}
\frac{\pd \bar{M}}{\pd t_k}=[(L^k)_+,\bar{M}],\quad
 \frac{\pd \bar{M}}{\pd \hat{t}_k}=[-(\hat{L}^k)_-, \bar{M}],\ \ k\in\Zop.
\end{equation}
\end{lem}

For an operator $A=A(L,\hat L^{-1},M,\hat M)$ which can be written as an antisymmetric form as $A=B-D^{-1} B^* D$, define a flow $Y_A$ acting on $\Phi$ and $\hat \Phi$ as
\eqa \label{genera}Y_A\Phi=-A_-\Phi,\ \ Y_A\hat \Phi=A_+\hat \Phi,\eeqa
therefore
\[Y_A L=[-A_-,L],\ \ Y_A\hat L=[A_+,\hat L],\]
and
\[Y_A M=[-A_-,M],\ \ Y_A\hat M=[A_+,\hat M].\]

Following calculation can be easily got
\[[Y_A,Y_B]\Phi= -(Y_AB)_- \Phi+ (Y_BA)_- \Phi+[B_-,A_-]\Phi,\]
\[[Y_A,Y_B]\hat\Phi= (Y_AB)_+ \hat\Phi-(Y_BA)_+ \hat\Phi+[B_+,A_+]\hat\Phi.\]
Above two identities can be written as an universal form
\eqa\label{twobrack}[Y_A,Y_B]\Phi(\hat \Phi)= Y_{\{B,A\}} \Phi(\hat \Phi),\eeqa
where
\eqa\{A,B\}=-Y_AB+Y_BA-[A_-,B_-]+[A_+,B_+].\eeqa
Also one can derive following proposition.
\begin{prp}\label{At}
For any polynomial $A=A(L,\hat L^{-1},M,\hat M)$, one has
\begin{align}
&\frac{\pd A}{\pd t_k}=[(L^k)_+, A], \quad \frac{\pd A}{\pd \hat t_k}=[-(\hat L^k)_-,A],\ \ k\in\Zop.
\end{align}
\end{prp}
\begin{prf}
Proof is easy to finish by considering eqs.\eqref{bkpLax} and eqs.\eqref{bkpMt}.
\end{prf}

Using eq.\eqref{genera} and Proposition \ref{At}, it can be proved
that the flow eqs.\eqref{genera} can commute with original flow of
the two-component BKP hierarchy, i.e.
\eqa\label{gensymm}[Y_A,\frac{\pd }{\pd  t_k}]=0,\ \ [Y_A,\frac{\pd }{\pd  \hat t_k}]=0.\eeqa That
means they are symmetries of the two-component BKP hierarchy. This
kind of symmetries contain original additional $w^B_{\infty}\times
w^B_{\infty}$ symmetry of the two-component BKP hierarchy mentioned
in \cite{bkpds}. The definition of operator $A$ here is more general
than operators used to construct additional symmetry of two
component BKP hierarchy in \cite{bkpds} because the multiplication
mixed set $\{L,M\}$ and set $\{\hat L,\hat M\}$ together. Therefore
we call it the generalized additional symmetry of the two-component
BKP hierarchy. Here we will not give a detailed proof of this
symmetry but later we will prove some special symmetry of this kind
of generalized additional symmetries.

The new bracket structure $\{\ , \ \}$ can be expressed by the standard bracket structure $[\  ,\  ]$
which is showed in the following lemma.
\begin{lem}
Following relations between two bracket structure hold
\eqa\{f\hat f,g\hat g\}=[f,g]\hat f\hat g-fg[\hat f,\hat g],\eeqa
\eqa\{\hat ff,\hat gg\}=\hat f\hat g[f,g]-[\hat f,\hat g]fg,\eeqa
\eqa\label{mixbra}\{f\hat f,\hat gg\}=[f,\hat g][g,\hat f]-f[\hat f,\hat g]g+\hat g[f,g]\hat f,\eeqa
where $f,g$ are polynomials of $L,M$ and $\hat f,\hat g$ are polynomials of $\hat L,\hat M.$
\end{lem}
\begin{prf}
The first two identities can be easily derived by direct calculation basing on definition, therefore we only give the proof of the identity \eqref{mixbra} as following
\begin{align*}
\{f\hat f,\hat gg\}&=-Y_{f\hat f}(\hat gg)+Y_{\hat gg}(f\hat f)+[(f\hat f)_+,(\hat gg)_+]-[(f\hat f)_-,(\hat gg)_-]\\
&=-[(f\hat f)_+,\hat g]g+\hat g[(f\hat f)_-, g]-[(\hat gg)_-,f]\hat f+f[(\hat gg)_+,\hat f]\\
& +[(f\hat f)_+,(\hat gg)_+]-[(f\hat f)_-,(\hat gg)_-]\\
&=f\hat gg\hat f+\hat g f\hat f g-f\hat f\hat gg-\hat g gf\hat f \\
&=(f\hat g-\hat g f)(g\hat f-\hat f g)-f\hat f\hat gg+f\hat g\hat f g+\hat gfg\hat f-\hat g gf\hat f \\
&=[f,\hat g][g,\hat f]-f[\hat f,\hat g]g+\hat g[f,g]\hat f.
\end{align*}
\end{prf}

From eq.\eqref{twobrack}, it is easy to see following lemma holds.
\begin{lem}
There is an antihomorphism between two sets, i.e.  $\C[L,\hat L,M,\hat M]$ and $\cG=\{Y_A|A=A(L,\hat L^{-1},M,\hat M)\}$
\begin{eqnarray*}
 \C[L,\hat L,M,\hat M],\{\ \ \}&\mapsto&, \cG, [\ \ ] ,\\
A&\mapsto&,Y_A,
\end{eqnarray*}
which satisfy following antihomorphism relation
\[[Y_A,Y_B]\Phi(\hat \Phi)= Y_{\{B,A\}} \Phi(\hat \Phi).\]
\end{lem}
Because of the anti-order of spectral representation of multiplications of Lax operators and Orlov-Schulman operators, following lemmas can be easily derived.
\begin{lem}
For $a_1,a_2,b_1,b_2 \in \Z_+$, there is an anti homorphism
\begin{eqnarray*}
\omega_{\infty}\otimes \omega_{\infty},[\ \ ]&\mapsto& \C[L,\hat L^{-},M,\hat M],\{\ \ \},\\
z_1^{a_1}\partial_{z_1}^{b_1}z_2^{a_2}\partial_{z_2}^{b_2}&\mapsto&M^{b_1} L^{a_1}\hat L^{-a_2}\hat M^{b_2},
\end{eqnarray*}
\[[z_1^{a_1}\partial_{z_1}^{b_1}z_2^{a_2}\partial_{z_2}^{b_2},z_1^{c_1}\partial_{z_1}^{d_1}z_2^{c_2}\partial_{z_2}^{d_2}]
\mapsto \{M^{d_1}L^{c_1}\hat L^{-d_2}\hat M^{d_2},M^{b_1}L^{a_1}\hat L^{-a_2}\hat M^{b_2}\}.\]
\end{lem}

\begin{lem}
For $a_1,a_2,b_1,b_2 \in \Z_+$, there is anisomorphism
\begin{eqnarray*}
\psi: \ \ \omega_{\infty}\otimes \omega_{\infty},[\ \ ]&\mapsto& \cG, [\ \ ] ,\\
z_1^{a_1}\partial_{z_1}^{b_1}z_2^{a_2}\partial_{z_2}^{b_2}&\mapsto& Y_{M^{b_1}L^{a_1}\hat L^{-a_2}\hat M^{b_2}},
\end{eqnarray*}
\[[z_1^{a_1}\partial_{z_1}^{b_1}z_2^{a_2}\partial_{z_2}^{b_2},z_1^{c_1}\partial_{z_1}^{d_1}z_2^{c_2}\partial_{z_2}^{d_2}]
\mapsto Y_{\{M^{d_1}L^{c_1}\hat L^{-d_2}\hat M^{d_2},M^{b_1}L^{a_1}\hat L^{-a_2}\hat M^{b_2}\}},\]
i.e.
\eqa[\psi(z_1^{a_1}\partial_{z_1}^{b_1}z_2^{a_2}\partial_{z_2}^{b_2}),\psi(z_1^{c_1}\partial_{z_1}^{c_1}z_2^{d_2}\partial_{z_2}^{d_2})]=
\psi([z_1^{a_1}\partial_{z_1}^{b_1}z_2^{a_2}\partial_{z_2}^{b_2},z_1^{c_1}\partial_{z_1}^{d_1}z_2^{d_2}\partial_{z_2}^{d_2}]).\eeqa
\end{lem}

From now on, we will introduce one special kind of case of $A=A(L,\hat L^{-1},M,\hat M)$, i.e. the following two operators $B_{m l}$ and $\hat{B}_{m l}$.
Given any pair of integers $(m,l)$ with $m,l\ge0$, define
\begin{align}\label{defBoperator}
B_{m l}=ML^{m+1}\hat L^{-l}+(-1)^{l+m}\hat L^{-l}L^{m}ML, \\
\ \ \hat{B}_{m l}=L^m\hat L^{-l+1}\hat M+(-1)^{l+m}\hat L\hat M\hat L^{-l}L^{m}.
\end{align}
The definitions of $B_{m l}$ and $\hat{B}_{m l}$ are also different from definitions in \cite{bkpds}.
As a corollary of Proposition \ref{At}, following proposition can be got.
\begin{prp}
For any $\bar B_{m l}=B_{m l}, \hat{B}_{m l}$, one has
\begin{align}\label{Bflow}
&\frac{\pd \bar B_{m l}}{\pd t_k}=[(L^k)_+, \bar B_{m l}], \quad \frac{\pd \bar B_{m l}}{\pd \hat t_k}=[-(\hat L^k)_-,\bar B_{m l}],\ \ k\in\Zop.
\end{align}
\end{prp}

To prove that $B_{m l}$ and $\hat B_{m l}$ satisfy B type condition, we need following lemma.
\begin{lem}\label{BtypM}
Operators $M$and $\hat M$ satisfy following conjugate identities,
\eqa
M^*
=DL^{-1}ML D^{-1},\ \ \ \hat M^*
=D\hat L\hat M\hat L^{-1} D^{-1}.\eeqa
\end{lem}
\begin{prf}
Using
\[
 \Phi^*=D\Phi^{-1} D^{-1},\ \ \hat\Phi^*=D\hat\Phi^{-1} D^{-1},\]
  following calculations
\[
M^* =\Phi^{*-1}\Gamma \Phi^*=D\Phi D^{-1}\Gamma D\Phi^{-1} D^{-1}
=D\Phi D^{-1}\Phi^{-1}M\Phi D\Phi^{-1} D^{-1},\]

\[
\hat M^* =\hat \Phi^{*-1}\hat \Gamma \hat \Phi^*=D\hat \Phi D^{-1}\hat \Gamma D\hat \Phi^{-1} D^{-1}
=D\hat \Phi D^{-1}\hat \Phi^{-1}\hat M\hat \Phi D\hat \Phi^{-1} D^{-1},\]
will lead to this lemma.
\end{prf}

It is easy to check following proposition holds basing on the Lemma \ref{BtypM} above.
\begin{prp}\label{asym}
$B_{m l}$ and $\hat B_{m l}$ satisfy B type condition, namely
\begin{equation}
B_{m l}^*=-D  B_{m l} D^{-1}, \quad \hat{B}_{m l}^*=-D \hat{B}_{m l}
 D^{-1}.
\end{equation}
\end{prp}
\begin{prf}
Using Proposition \ref{BtypM}, following calculation will lead to first identity of this proposition
\begin{eqnarray*}B_{m l}^*
&=&(ML^{m+1}\hat L^{-l}-(-1)^{l+m+1}\hat L^{-l}L^{m}M L)^*\\
&=&\hat L^{-l*}L^{m+1*}M^*-(-1)^{l+m+1}L^{1*}M^*L^{m*}\hat L^{-l*}\\
&=&(-1)^{l+m+1}D\hat L^{-l}L^{m}MLD^{-1}-DML^{m+1}\hat L^{-l}D^{-1}\\
&=&-D(ML^{m+1}\hat L^{-l}-(-1)^{l+m+1}\hat L^{-l}L^{m}M L)D^{-1}.
\end{eqnarray*}
 The second identity can be proved in similar way.
\end{prf}

Because of Proposition \ref{asym}, the following equations are well defined
\begin{align}
&\frac{\pd \Phi}{\pd b_{m l}}=Y_{B_{m l}}\Phi=- (B_{m l})_-\Phi,
\quad \frac{\pd \hat{\Phi}}{\pd b_{m l}}=Y_{B_{m l}}\hat\Phi=(B_{m l})_+\hat{\Phi}, \label{add}\\
&\frac{\pd \Phi}{\pd \hat{b}_{m l}}=Y_{\hat B_{m l}}\Phi=- (\hat{B}_{m l})_-\Phi, \quad
\frac{\pd \hat{\Phi}}{\pd \hat{b}_{m l}}=Y_{\hat B_{m l}}\hat\Phi=(\hat{B}_{m
l})_+\hat{\Phi}. \label{add2}
\end{align}
These equations are equivalent to following Lax equations
\begin{align}
&\frac{\pd L}{\pd b_{m l}}=[- (B_{m l})_-,L],
\quad \frac{\pd \hat{L}}{\pd b_{m l}}=[(B_{m l})_+,\hat L], \label{add}\\
&\frac{\pd L}{\pd \hat{b}_{m l}}=[- (\hat{B}_{m l})_-,L], \quad
\frac{\pd \hat{L}}{\pd \hat{b}_{m l}}=[(\hat{B}_{m
l})_+,\hat{L}]. \label{add2}
\end{align}
These flows are in fact some special cases of generalized additional symmetries eqs.\eqref{genera}. To show some techniques in the proof of generalized symmetry eq.\eqref{gensymm}, we will give a short proof of following proposition.
\begin{prp}\label{thm-st}
The flows \eqref{add} and \eqref{add2} commute with the flows of  the  two-component BKP hierarchy. Namely, for any $\bar{b}_{m l}=b_{m l}, \hat{b}_{m l}$
and $\bar{t}_k=t_k, \hat{t}_k$ one has
\begin{equation}\label{st}
\left[\frac{\pd}{\pd \bar{b}_{m l}}, \frac{\pd}{\pd
\bar{t}_k}\right]=0, \quad m,l\in\Z_+, ~~ k\in\Zop,
\end{equation}
which holds in the sense of acting on  $\Phi$ or $\hat\Phi$.
\end{prp}
\begin{prf}
The  proposition can be checked case by case with the help of
eq.\eqref{Bflow} and eqs.\eqref{add}-\eqref{add2}.
For example,

\begin{align}\label{}
&\left[\frac{\pd}{\pd {b}_{m l}}, \frac{\pd}{\pd
\hat t_k}\right]\Phi \nn\\
=& [ (\hat L^k)_-,({B}_{m l})_-]\Phi
-[({B}_{m l})_+, \hat L^k]_- \Phi -[(\hat L^k)_-,{B}_{m l}]_-\Phi =0, \nn
\end{align}
\begin{align}\label{}
&\left[\frac{\pd}{\pd \hat{b}_{m l}}, \frac{\pd}{\pd
t_k}\right]\hat{\Phi} \nn\\
=& [ (L^k)_+-\dt_{k1}\hat{L}^{-1},(\hat{B}_{m l})_+]\hat{\Phi}
+\left([-(\hat{B}_{m l})_-, L^k]_+ - \dt_{k1}[(\hat{B}_{m l})_+,
\hat{L}^{-1}]\right)\hat{\Phi} \nn \\
&-[(L^k)_+,\hat{B}_{m l}]_+\hat{\Phi} =0. \nn
\end{align}
The other cases can be proved in similar ways. This is the end of this proposition.
\end{prf}

This proposition implies that the additional flows
\eqref{add}-\eqref{add2} are symmetries of the  two-component BKP
hierarchy. To see the further structure of the additional symmetry,
we need following proposition.
\begin{prp}
The operators $\bar B_{m,n}=B_{m,n},\hat B_{m,n}, m,n\in \Z_+$ of the  two-component BKP hierarchy satisfy following identity
\begin{align*}
&\{B_{m_1,m_2},B_{n_1,n_2}\}=(m_1-n_1)B_{m_1+n_1,m_2+n_2}+Q_{m_1,m_2,n_1,n_2}, \\
&
\{\hat B_{m_1,m_2},\hat B_{n_1,n_2}\}=(m_2-n_2)\hat B_{m_1+n_1,m_2+n_2}+\hat Q_{m_1,m_2,n_1,n_2},
\\
&\{B_{m_1,m_2},\hat B_{n_1,n_2}\}=-n_1\hat B_{m_1+n_1,m_2+n_2}+m_2 B_{m_1+n_1,m_2+n_2}+\bar Q_{m_1,m_2,n_1,n_2},
\end{align*}
 where

\begin{eqnarray*}Q_{m_1,m_2,n_1,n_2}&=&(-1)^{m_2+m_1+1}\{ML^{n_1+1}\hat L^{-n_2},\hat L^{-m_2}L^{m_1}M L\}\\
&&+(-1)^{n_2+n_1}\{ML^{m_1+1}\hat L^{-m_2},\hat L^{-n_2}L^{n_1}M L\},
\end{eqnarray*}

\begin{eqnarray*}\hat Q_{m_1,m_2,n_1,n_2}=&&(-1)^{m_2+m_1+1}\{ L^{n_1}\hat L^{-n_2+1}\hat M,\hat L\hat M\hat L^{-m_2}L^{m_1}\}\\
&&+(-1)^{n_2+n_1}\{L^{m_1}\hat L^{-m_2+1}\hat M,\hat L\hat M\hat L^{-n_2}L^{n_1}\},
\end{eqnarray*}

\begin{eqnarray*}\bar Q_{m_1,m_2,n_1,n_2}=&&(-1)^{m_2+m_1+1}\{ L^{n_1}\hat L^{-n_2+1}\hat M,\hat L^{-m_2}L^{m_1}M L\}\\
&&+(-1)^{n_2+n_1}\{ML^{m_1+1}\hat L^{-m_2},\hat L\hat M\hat L^{-n_2}L^{n_1}\}.\end{eqnarray*}
\end{prp}
Take
\[\tilde B_{m_1,m_2}=(m_2+1)B_{ m_1,m_2}-m_1\hat B_{m_1,m_2},\]
then
\[\{\tilde B_{m_1,m_2},\tilde B_{n_1,n_2}\}=((n_2+1)m_1-(m_2+1)n_1)\tilde B_{m_1+n_1,m_2+n_2}+S_{m_1,m_2,n_1,n_2},\]
where
\begin{eqnarray*}&&S_{m_1,m_2,n_1,n_2}=(m_2+1)(n_2+1)Q_{m_1,m_2,n_1,n_2}-n_1(m_2+1)\bar Q_{m_1,m_2,n_1,n_2}\\
&&+m_1(n_2+1)\bar Q_{n_1,n_2,m_1,m_2}+m_1n_1\hat Q_{m_1,m_2,n_1,n_2}.
\end{eqnarray*}

Then the following theorem is clear.
\begin{thm}\label{modblock}
In the sense of acting on  $\Phi$ or $\hat\Phi$, the additional flows \eqref{add} and \eqref{add2} satisfy following relations
\begin{align}
&[\pd_{b_{m_1,m_2}},\pd_{b_{n_1,n_2}}]=(m_1-n_1)\pd_{b_{m_1+n_1,m_2+n_2}}+Y_{Q_{m_1,m_2,n_1,n_2}}, \\
&
[\pd_{\hat b_{m_1,m_2}},\pd_{\hat b_{n_1,n_2}}]=(m_2-n_2)\pd_{\hat b_{m_1+n_1,m_2+n_2}}+Y_{\hat Q_{m_1,m_2,n_1,n_2}},
\\
&[\pd_{ b_{m_1,m_2}},\pd_{\hat b_{n_1,n_2}}]=m_2\pd_{ b_{m_1+n_1,m_2+n_2}}-n_1\pd_{\hat b_{m_1+n_1,m_2+n_2}}+Y_{\bar Q_{m_1,m_2,n_1,n_2}}.\end{align}
These above relations further lead to following modified Block type algebraic relation
\begin{align}
[\pd_{v_{m_1,m_2}},\pd_{v_{n_1,n_2}}]=((n_2+1)m_1-(m_2+1)n_1)\pd_{ v_{m_1+n_1,m_2+n_2}}+Y_{S_{m_1,m_2,n_1,n_2}},
\end{align}
where
\[\pd_{v_{m_1,m_2}}=(m_2+1)\pd_{ b_{m_1,m_2}}-m_1\pd_{\hat b_{m_1,m_2}}.\]

\end{thm}

Without B type condition eq.\eqref{Bcondition}, the operators
$(Q,\hat Q,S,Y_{Q},Y_{\hat Q},Y_{S})$ will vanish. This will lead to
nice Block symmetric structure. Proposition~\ref{thm-st} and Theorem
\ref{modblock} show that  the flows \eqref{add} and \eqref{add2}
give one modified Block type additional symmetries for the
two-component BKP hierarchy. The obstacle to derive the perfect
Block type symmetry is due to the constrained B type condition
eq.\eqref{Bcondition} of the two-component BKP hierarchy. That means
if we only consider the the two-component KP hierarchy, i.e. the
two-component BKP hierarchy without the constrained B type
condition, the additional symmetry will compose nice  structure of
Block type infinite dimensional Lie algebra.  But unfortunately the
Lax representation with two different pseudo-differential operators
of the two-component KP hierarchy is not well-defined.

If we choose  $l=0$ in the operator $ B_{m,l}$, $m=0$ in the operator $\hat B_{m,l} $ and increase one index on each operator of $M,\hat M$, then this symmetry will be the $w^B_{\infty}\times w^B_{\infty}$ algebra
mentioned in \cite{bkpds}.

Although we only get modified Block type additional symmetries for the two-component BKP
 hierarchy, further calculation in the next section supports:
 If we do a (2n,2)-reduction from the two-component BKP hierarchy,
 perfect Block type additional  symmetry will  be exactly kept.
This reduced hierarchy is nothing but the D type Drinfeld--Sokolov
hierarchies \cite{DS} which will be discussed in the next section.

\section{D type Drinfeld--Sokolov hierarchy}

Assume a new Lax operator $\cL$ which has following relation with two Lax operators of the two-component BKP hierarchy introduced in last section
\begin{equation}\label{constraint}\cL=L^{2n}=\hat{L}^{2},\ n\geq 2.\end{equation}
Then the Lax operators of two-component BKP hierarchy will be reduced to the following Lax operator of D type Drinfeld--Sokolov hierarchy\cite{LWZ,bkpds}
\begin{equation}\label{mL}
   \cL=D^{2n}+\frac1{2}\sum_{i=1}^{n} D^{-1}\left(v_i
D^{2i-1}+D^{2i-1} v_i\right) +D^{-1} \rho D^{-1} \rho.
\end{equation}
The difference of the  Lax operator $\cL$ from the one of the D type Drinfeld--Sokolov hierarchy in \cite{LWZ,bkpds} is we did a shift on $n$, i.e. we change $n-1,n\geq 3$ to $n,n\geq 2$ for simplicity. That will not affect the system itself at all.

{\bf Remark:}
It seems that one can not compute the square of the operator $\hat L$, because  it contains infinite terms  with positive powers of $D$ and is not a pseudo-differential operator in common sense.
In eqs.(6.1)-(6.2) in \cite{bkpds}, Chaozhong Wu give the above reduction directly without a proof because in paper \cite{LWZ} they have spent a lot of space to carry out the proof. Here we only describe some key points of the proof in \cite{LWZ} which is in a inverse direction.
The Lemma 3.1 and Lemma 3.3 in \cite{LWZ} tell us, for a given operator $ \cL$ in eq.\eqref{mL}, there exists two fractional operators $\cL^{\frac1{2n}}$ and $\cL^{\frac1{2}}$ in same forms as $L$ and $\hat L$ (eq.\eqref{PhP}) in different operator rings. In \cite{LWZ}, they prove that one can choose two fractional operators to be exactly the Lax operators $L$ and $\hat L$ in the two-component BKP hierarchy in last section because they satisfy all the necessary characters such as antisymmetric property. The difficulty is in the proof of Lemma 3.3 in \cite{LWZ} with the help of Lemma 2.5 in \cite{LWZ} which promise the reasonability to define the square root of the pseudo-differential operator $\cL$.
Therefore to save the space, we will not give the repeated proof as \cite{LWZ} on the consistency between the  D type Drinfeld--Sokolov hierarchy and the two-component BKP hierarchy under the reduction condition eq.\eqref{constraint}.

One can easily find the Lax operator $\cL$ of  D type Drinfeld--Sokolov hierarchy will not satisfy the reduction condition as Lax operator of the two-component BKP hierarchy but satisfy following B type condition
\eqa\label{symcL}\cL^*=D \cL D^{-1}.\eeqa
This Lax operator $\cL$ of D type Drinfeld--Sokolov hierarchy has following dressing structure\cite{bkpds}
\begin{equation} \label{dress}
\cL=\Phi D^{2n}\Phi^{-1}=\hat{\Phi} D^{-2}\hat{\Phi}^{-1}.
\end{equation}
Here
\begin{align} \label{Phi}
\Phi=1+\sum_{i\ge 1}a_i D^{-i},\quad \hat{\Phi}=1+\sum_{i\ge 1}b_i
D^{i}
\end{align}
are pseudo-differential operators  that also
satisfy following B type condition
\begin{equation}\label{phipsi}
\Phi^*= D\Phi^{-1} D^{-1},\quad \hat{\Phi}^*= D\hat{\Phi}^{-1}
D^{-1}.
\end{equation}
The dressing structure inspire us to define two fractional operators as
\begin{equation}
\cL^{\frac1{2n}}= D+\sum_{i\ge1}u_i  D^{-i}, \quad \cL^{\frac12}=
D^{-1}\hat{u}_{-1}+\sum_{i\ge1}\hat{u}_i D^i.
\end{equation}

Two fractional operators $\cL^{\frac1{2n}}$ and $\cL^{\frac12}$ can be rewritten  in a dressing form as
\begin{equation} \label{PPh}
\cL^{\frac1{2n}}=\Phi D\Phi^{-1},\quad \cL^{\frac12}=\hat{\Phi} D^{-1}\hat{\Phi}^{-1}.
\end{equation}

 The   D type Drinfeld--Sokolov hierarchy being considered in this paper is defined by the following
Lax equations:
\begin{align}\label{PPht}
& \frac{\pd \cL}{\pd t_k}=[(\cL^{\frac k{2n}})_+, \cL], \quad \frac{\pd \cL}{\pd \hat t_k}=[-(\cL^{\frac k2})_-, \cL],\ \ k\in\Zop.
\end{align}
Among these hierarchies, the Drinfeld--Sokolov hierarchy of type
$D_n$  is associated to the affine algebra $D_n^{(1)}$ and the
zeroth vertex of its Dynkin diagram \cite{DS,LWZ}.
Similarly as the two-component BKP hierarchy, the equivalence between  $\pd/\pd t_1$ and $\pd/\pd x$ leads to assumption as $t_1=x$.

The dressing operators $\Phi$ and $\hat \Phi$ are same as the ones of two-component BKP hierarchy.
Given $\cL$, the dressing operators $\Phi$ and
$\hat{\Phi}$ are  uniquely determined up to a multiplication to the
right by operators of the form \eqref{Phi} and \eqref{phipsi} with
constant coefficients. The D type Drinfeld-Sokolov hierarchies
can also be redefined as
\begin{align}
&\frac{\pd \Phi}{\pd t_k}=- (\cL^{\frac k{2n}})_-\Phi, \quad
\frac{\pd \hat{\Phi}}{\pd t_k}=\bigl((\cL^{\frac k{2n}})_+ -\dt_{k1} \cL^{-\frac1{2}}\bigr)\hat{\Phi}, \label{ppt1}\\
&\frac{\pd \Phi}{\pd \hat{t}_k}=- (\cL^{\frac k{2}})_-\Phi, \quad \frac{\pd
\hat{\Phi}}{\pd \hat{t}_k}=(\cL^{\frac k{2}})_+\hat{\Phi} \label{ppt2}
\end{align}
with $k\in\Zop$.

Introduce
two wave functions
\begin{align}\label{wavef}
&w(z^{\frac 1{2n}})=w(t, \hat{t}; z^{\frac 1{2n}})=\Phi e^{\xi(t; z^{\frac 1{2n}})},
\\
& \hat{w}(z^{\frac 12})=\hat{w}(t, \hat{t}; z^{\frac 1{2}})=\hat{\Phi}
e^{x  z^{\frac 1{2}}+\xi(\hat{t};-z^{-\frac 1{2}})}.
\end{align}
 It is easy to see
\[
\cL\,w(z^{\frac 1{2n}})=z w(z^{\frac 1{2n}}), \quad \cL \hat{w}(z^{\frac 12}) = z^{-1} \hat{w}(z^{\frac 12}).
\]

After above preparation, we will show that this D type Drinfeld-Sokolov hierarchies have nice Block symmetry as its appearance  in BTH
\cite{ourBlock}.

\section{Block symmetries of D type Drinfeld-Sokolov hierarchies }

In this section, we will put constrained condition
eq.\eqref{constraint} into construction of the flows of additional
symmetry which form the well-known Block algebra.

With the dressing operators given in eq.\eqref{PPh}, we introduce Orlov-Schulman operators as following
\begin{equation*}
\cM=\Phi\Gm_{L}\Phi^{-1}, \quad
\hat{\cM}=\hat{\Phi}\hat{\Gm}_{R}\hat{\Phi}^{-1},
\end{equation*}
where
\[
\Gm_{L}=\sum_{k\in\Zop}\frac{k}{2n} t_k  D^{k-2n}, \quad \hat{\Gm}_{R}=-\frac x{2}D^{3}-\frac1{2}\sum_{k\in\Zop}
k\hat{t}_k D^{2-k}.
\]

It is easy to see the following lemma holds.
\begin{lem}\label{thm-Mw}
The operators $\cM$ and $\hat{\cM}$ satisfy
\begin{equation}\label{}
[\cL, \cM]=1, \quad [\cL,\hat{\cM}]=1;
\end{equation}
and
\begin{equation}\label{}
\cM w(z^{\frac 1{2n}})=\pd_{z} w(z^{\frac 1{2n}}), \quad \hat{\cM} \hat{w}(z^{\frac 12})=\pd_{z^{-1}} \hat{w}(z^{\frac 12});
\end{equation}
\begin{equation}\label{Mt}
\frac{\pd \bar\cM}{\pd t_k}=[(\cL^{\frac k{2n}})_+,\bar\cM],\quad
 \frac{\pd \bar\cM}{\pd \hat{t}_k}=[-(\cL^{\frac k{2}})_-, \bar\cM],
\end{equation}
where
 $\bar{\cM}=\cM$ or $\hat{\cM}, k\in\Zop$.
\end{lem}

To make the operators used in additional symmetry satisfying B type condition, we need to prove the following $B$ type property of $\cM-\hat \cM$ which is
included in following lemma.

\begin{lem}\label{asymM-M2}
The difference of two Orlov-Schulman operators $\cM$ and $\hat \cM$ for D type Drinfeld-Sokolov hierarchy has following D type property:
\begin{align}
\cL^*(\cM-\hat \cM)^*=-D\cL(\cM -\hat\cM )D^{-1}.
\end{align}
\end{lem}
\begin{prf}
It is easy to find  the two Orlov-Schulman operators $\cM$ and $\hat \cM$ of the D type Drinfeld-Sokolov hierarchy can be expressed by Orlov-Schulman operators $M,\hat M$ and Lax operators $L,\hat L$ of two-component BKP hierarchy as
\eqa \label{cMandM}\cM=\frac{ML^{1-2n}}{2n},\ \ \hat \cM=-\frac{\hat M\hat L^{-3}}{2}.\eeqa

Using Lemma \ref{BtypM}, putting eq.\eqref{cMandM} into $(\cM-\hat \cM)^*$ can lead to
\begin{align}
&(\cM-\hat \cM)^*=-\frac{DL^{-2n}MLD^{-1}}{2n}-\frac{D\hat L^{-2}\hat M\hat L^{-1} D^{-1}}{2}\\
&=-\frac{DL^{-2n}MLD^{-1}}{2n}-\frac{D\hat L^{-2}\hat M\hat L^{-1} D^{-1}}{2}\\
&=-\frac{D(ML^{1-2n}-2nL^{-2n})D^{-1}}{2n}-\frac{D(\hat M\hat L^{-3}+2\hat L^{-2}) D^{-1}}{2},
\end{align}
which can further lead to
\begin{align}
\cL^*(\cM-\hat \cM)^*=-D(\cL\cM -\cL\hat\cM)D^{-1}.
\end{align}

In above calculation, the commutativity between $\cL$ and $\cM -\hat\cM$ is already used.
Till now, the proof is finished.
\end{prf}

For D-type Drinfeld-Sokolov hierarchy, $m$ is supposed to be odd number to avoid being trivial and simplify it  to
\begin{align}
\B_{m,l}=(\cM-\hat \cM)^{m}\cL^l,\ \  m\in \Zop.
\end{align}
One can easily check that
\begin{align}
\B_{m,l}^*=-D\B_{m,l}D^{-1},\ \  m\in \Zop.
\end{align}
That means it is reasonable to define additional flow of the D type Drinfeld--Sokolov hierarchy
\begin{align}\label{blockflow}
&\frac{\pd \cL}{\pd c_{m,l}}=[-({\B}_{m,l})_-, \cL],\ \ m\in \Zop, l\in \Z_+.
\end{align}

\begin{prp}
For the Drinfeld--Sokolov hierarchy of type $D$, the flows \eqref{blockflow} can commute with original flow of  the Drinfeld--Sokolov hierarchy of type $D$, namely,
\begin{equation*}
\left[\frac{\pd}{\pd c_{m,l}}, \frac{\pd}{\pd t_k}\right]=0, \quad
\left[\frac{\pd}{\pd c_{m,l}}, \frac{\pd}{\pd\hat{t}_k}\right]=0, \qquad l\in \Z_{+}, ~m,k\in\Zop,
\end{equation*}
which hold in the sense of acting on  $\Phi$, $\hat\Phi$ or $\cL.$

\end{prp}
\begin{prf} According to the definition,
\begin{eqnarray*}
[\partial_{c_{m,l}},\partial_{t_k}]\Phi=\partial_{c_{m,l}}
(\partial_{t_k}\Phi)-
\partial_{t_k} (\partial_{c_{m,l}}\Phi),
\end{eqnarray*}
and using the actions of the additional flows and the
flows of D type  Drinfeld-Sokolov hierarchy on $\Phi$,  we have
\begin{eqnarray*}
[\partial_{c_{m,l}},\partial_{t_k}]\Phi
&=& -\partial_{c_{m,l}}\left((\cL^{\frac{k}{2n}})_{-}\Phi\right)+
\partial_{t_k} \left(((\cM-\hat \cM)^m\cL^l)_{-}\Phi \right)\\
&=& -(\partial_{c_{m,l}}\cL^{\frac{k}{2n}} )_{-}\Phi-
(\cL^{\frac{k}{2n}})_{-}(\partial_{c_{m,l}}\Phi)\\&&+
[\partial_{t_k} ((\cM-\hat \cM)^m\cL^l)]_{-}\Phi +
((\cM-\hat \cM)^m\cL^l)_{-}(\partial_{t_k}\Phi).
\end{eqnarray*}
Using eq.\eqref{PPht} and eq.\eqref{Mt}, it
equals
\begin{eqnarray*}
[\partial_{c_{m,l}},\partial_{t_k}]\Phi
&=&[\left((\cM-\hat \cM)^m\cL^l\right)_{-}, \cL^{\frac{k}{2n}}]_{-}\Phi+
(\cL^{\frac{k}{2n}})_{-}\left((\cM-\hat \cM)^m\cL^l\right)_{-}\Phi\\
&&+[(\cL^{\frac{k}{2n}})_{+},(\cM-\hat \cM)^m\cL^l]_{-}\Phi-((\cM-\hat \cM)^m\cL^l)_{-}(\cL^{\frac{k}{2n}})_{-}\Phi\\
&=&[((\cM-\hat \cM)^m\cL^l)_{-}, \cL^{\frac{k}{2n}}]_{-}\Phi- [(\cM-\hat \cM)^m\cL^l,
(\cL^{\frac{k}{2n}})_{+}]_{-}\Phi\\&&+
[(\cL^{\frac{k}{2n}})_{-},((\cM-\hat \cM)^m\cL^l)_{-}]\Phi\\
&=&0.
\end{eqnarray*}
The other cases of this proposition can be proved in similar ways.
\end{prf}

Above proposition indicate that eq.\eqref{blockflow} is symmetry of D type Drinfeld-Sokolov hierarchy.
Further we can get following identities hold
\begin{equation}\label{blockflowM}
\frac{\pd \cM}{\pd c_{m,l}}=[-(\B_{m,l})_-,\cM], \quad
\frac{\pd\hat{\cM}}{\pd c_{m,l}}=[(\B_{m,l})_+,\hat{\cM}], \ \ m\in \Zop, l\in \Z_+,
\end{equation}
\begin{equation}\label{waves}
\frac{\pd w(z^{\frac 1{2n}})}{\pd c_{m,l}}=-(\B_{m,l})_-w(z^{\frac 1{2n}}), \quad
\frac{\pd\hat{w}(z^{\frac 12})}{\pd c_{m,l}}=(\B_{m,l})_+\hat{w}(z^{\frac 12}), \qquad j\ge-1.
\end{equation}

Using same technique used in \cite{ourBlock}, following theorem can be derived.
\begin{thm}\label{thm-MLs}
The flows in eq.\eqref{blockflow} about  additional symmetries of D type Drinfeld-Sokolov hierarchy compose following Block type Lie algebra

\[[\partial_{c_{m,l}},\partial_{c_{s,k}}]=(km-s l)\partial_{c_{m+s-1,k+l-1}},\ \  m,s\in \Zop, k,l\in \Z_+,\]

which holds in the sense of acting on  $\Phi$, $\hat\Phi$ or $\cL.$
\end{thm}
\begin{prf}
 By using eq.\eqref{blockflow} and eq.\eqref{blockflowM}, we get
\begin{eqnarray*}
[\partial_{c_{m,l}},\partial_{c_{s,k}}]\Phi&=&
\partial_{c_{m,l}}(\partial_{c_{s,k}}\Phi)-
\partial_{c_{s,k}}(\partial_{c_{m,l}}\Phi)\\
&=&-\partial_{c_{m,l}}\left(((\cM-\hat \cM)^s\cL^k)_{-}\Phi\right)
+\partial_{c_{s,k}}\left(((\cM-\hat \cM)^m\cL^l)_{-}\Phi\right)\\
&=&-(\partial_{c_{m,l}}
(\cM-\hat \cM)^s\cL^k)_{-}\Phi-((\cM-\hat \cM)^s\cL^k)_{-}(\partial_{c_{m,l}} \Phi)\\
&&+ (\partial_{c_{s,k}} (\cM-\hat \cM)^m\cL^l)_{-}\Phi+
((\cM-\hat \cM)^m\cL^l)_{-}(\partial_{c_{s,k}} \Phi),
\end{eqnarray*}
which further leads to
 \begin{eqnarray*}&&
[\partial_{c_{m,l}},\partial_{c_{s,k}}]\Phi\\
&=&-\Big[\sum_{p=0}^{s-1}
(\cM-\hat \cM)^p(\partial_{c_{m,l}}(\cM-\hat \cM))(\cM-\hat \cM)^{s-p-1}\cL^k
+(\cM-\hat \cM)^s(\partial_{c_{m,l}}\cL^k)\Big]_{-}\Phi\\&&-((\cM-\hat \cM)^s\cL^k)_{-}(\partial_{c_{m,l}} \Phi)\\
&&+\Big[\sum_{p=0}^{m-1}
(\cM-\hat \cM)^p(\partial_{c_{s,k}}(\cM-\hat \cM))(\cM-\hat \cM)^{m-p-1}\cL^l
+(\cM-\hat \cM)^m(\partial_{c_{s,k}}\cL^l)\Big]_{-}\Phi\\&&+
((\cM-\hat \cM)^m\cL^l)_{-}(\partial_{c_{s,k}} \Phi)\\
&=&[(s l-km)(\cM-\hat \cM)^{m+s-1}\cL^{k+l-1}]_-\Phi\\
&=&(km-s l)\partial_{c_{m+s-1,k+l-1}}\Phi.
\end{eqnarray*}
In the process of deriving the above nice algebraic structure, we omitted a lot of tedious calculation among operators.
Similarly  the same results on $\hat \Phi$ and $\cL$ can be got.

\end{prf}

Our early papers and above results show the Block type algebras are
appeared not only in Toda type difference systems
  but also in  differential systems such as two-BKP hierarchy, D type
  Drinfeld-Sokolov hierarchy, which represents one kind of  hidden symmetry
  algebraic structures of them. These results also show that
  Block infinite dimensional Lie algebra has a certain of
  universality in integrable hierarchies.

\vskip 0.5truecm \noindent{\bf Acknowledgments.}
We are grateful to Prof. Dafeng Zuo, Jipeng Cheng, Zhiwei Wu and Kelei Tian
for valuable discussions.  We also thank the referee for his/her
valuable suggestions on the proof of Theorem \ref{thm-MLs}. Chuanzhong Li is supported by the National Natural Science Foundation of China under Grant No. 11201251, the Natural Science Foundation of Zhejiang Province under Grant No. LY12A01007, the Natural Science Foundation of Ningbo under Grant No. 2013A610105. Jingsong He is supported by the National Natural Science Foundation of China under Grant No. 11271210, K.C.Wong Magna Fund in
Ningbo University.

\end{document}